Research article

# Integrating spectrophotometric and XRD analyses in the investigation of burned dental remains

Author names and affiliations.

**Rabiah A. Rahmat** [a,e*], **Melissa Humphries** [b], **Jeremy J. Austin** [a], **Adrian M. T. Linacre** [c], **Mark Raven** [d], **Peter Self** [d]

[a] *Australian Centre for Ancient DNA, School of Biological Sciences, University of Adelaide, Adelaide, South Australia 5006 Australia*
[b] *Faculty of Engineering, Computer and Mathematical Sciences, University of Adelaide, Adelaide, South Australia 5006, Australia*
[c] *College of Science and Engineering, Flinders University, Adelaide, South Australia 5042 Australia*
[d] *CSIRO, Land and Water, Locked Bag 2, Glen Osmond, South Australia 5064, Australia*
[e] *Department of Oro-Maxillofacial Surgical and Medical Sciences, Faculty of Dentistry, University of Malaya, Kuala Lumpur 50603, Malaysia*

Email addresses:

**Melissa Humphries**

melissa.humphries@adelaide.edu.au

**Jeremy J. Austin**

jeremy.austin@adelaide.edu.au

**Adrian M. T. Linacre**

adrian.linacre@flinders.edu.au

**Mark Raven**

Mark.Raven@csiro.au

**Peter Self**

Peter.Self@csiro.au

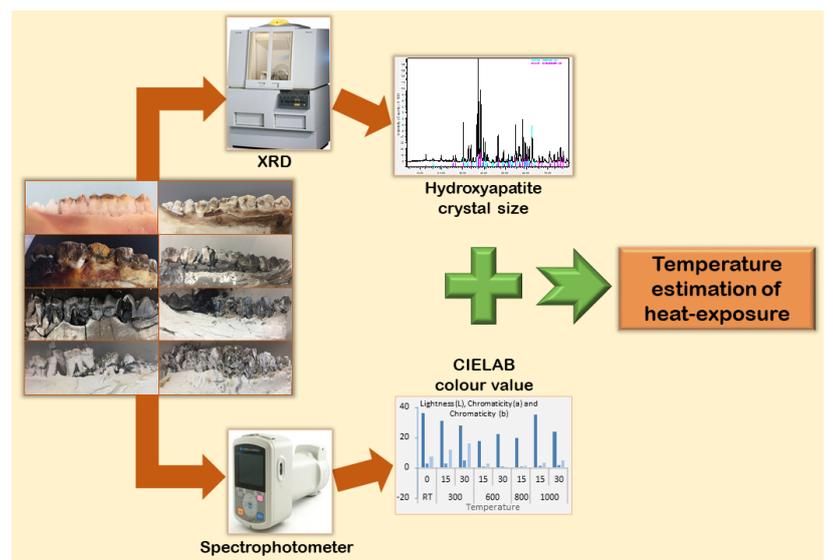

*Corresponding author. Australian Centre for Ancient DNA, School of Biological Sciences, University of Adelaide, Adelaide, South Australia 5006 Australia.
Permanent address: Department of Oro-Maxillofacial Surgical and Medical Sciences, Faculty of Dentistry, University of Malaya, Kuala Lumpur 50603, Malaysia
*E-mail address:* drrabiahadaw@gmail.com (R. A. Rahmat).

**Abstract**

Heat alters colour and crystallinity of teeth by destruction of the organic content and inducing hydroxyapatite crystal growth. The colour and crystallite changes can be quantified using spectrophotometric and x-ray diffraction analyses, however these analyses are not commonly used in combination to evaluate burned dental remains. In this study, thirty-nine teeth were incinerated at 300 – 1000°C for 15 and 30 minutes and then measured using a spectrophotometer and an x-ray diffractometer. Response variables used were lightness, $L^*$, and chromaticity $a^*$ and $b^*$ and luminance (whiteness and yellowness) for colour, and crystal size for crystallinity. Statistical analysis to determine the attribution of these variables revealed yellowness and crystal size were significantly affected by temperature ($p < 0.05$), whilst duration of heat-exposure showed no significant effect. This study suggests the inclusion of both spectrophotometric and x-ray diffraction in investigating thermal-heated teeth is useful to accurately estimate the temperature teeth are exposed to.

1. Introduction

The resilient structure of teeth to withstand the test of time and adverse environmental conditions means that teeth are frequently recovered from archaeological settings. Through burned dental remains, archaeology experts can analyse and interpret the demography, cultural practice and past ritual ceremony. Teeth also can be crucial evidence for human identification after a fire. Analysis of temperature-dependent characteristics of teeth including the alterations of colour and crystal size can provide contextual information about a fire's condition such as fire temperature, and eventually facilitates an investigation of burned dental remains (1-4). For this reason, the investigation of burned dental remains is of ongoing interest in archaeological and forensic science. Teeth, being the hardest structure of the human body and the heat insulation gained from the surrounding musculoskeletal structures, usually survive high temperatures and are the least of all body parts affected by the fire (5, 6). Burned dental remains found in archaeological sites (e.g. cultural cremation practice), and forensic cases (e.g. aircraft accidents, vehicle and house fires, and bushfire) are commonly fragmented (7). Fragmented tooth crowns are usually recovered at a fire scene because they tend to break and fall apart due to direct heat exposure (8). These dental fragments are suitable to be used for spectrophotometric and x-ray diffraction (XRD) analyses.

Teeth are made up of inorganic and organic components in various proportions. For example: enamel is composed of 97 wt.% of inorganic matter (hydroxyapatite) and 3 wt.% of organic matter (collagen, proteins and lipids); and dentin is composed of 70 wt.% of hydroxyapatite and 30% of organic matter (9). The organic matter in teeth is present in an aqueous-gel environment where collagen, proteins and lipids are kept hydrated (10). On heating, teeth undergo extensive structural changes and the ratio of mineral-organic contents are gradually altered. From 110 – 260°C, dehydration occurs in which hydroxyl bonds break and eventually teeth lose any water molecules (11). During this time, collagen gradually degrades (12). When the temperature is increased up to 500°C, combustion of the organic matter begins to occur especially in collagen-rich dentin (13, 14). The depletion of organic matter decreases the pH inside the tooth and leaves voids allowing the growth of hydroxyapatite (HA) crystals to fill up the interspaces further reducing the environmental pH to less than pH 4.2 (15). HA is at its the most stable at neutral pH, therefore in an acidic environment HA crystals become unstable (16). With an excess of $Mg^{2+}$ ions, the growth of HA is halted and HA transformed into a more stable form of calcium phosphate known as whitlockite [$Ca_9Mg(HPO_4)(PO_4)_6$] (17). Above 1100°C, HA crystals melt and coalesce to each other (18). This process can be reflected through alterations in the outward appearance of teeth including colour and crystallinity (18).

*1.1. Colour analysis*

The colour of teeth is said to be influenced by factors including lighting conditions, translucency, opacity, light scattering and gloss (19). As the intensity of heating increases, tooth colour progressively changes. In general, the sequence of colour changes is from its neutral colour (yellowish white), to brown, black, blue-grey, and finally chalky white (8, 20, 21). Factors affecting these changes are temperature, duration, oxygen availability and other related influences (10, 22). Traditionally, many studies have analysed colour changes of incinerated teeth visually and compared the observation to a colour guide such as the Munsell colour system (1, 2, 10, 20, 21). Despite the claim made by Shipman *et al.* (10) that the Munsell colour chart system offers a standardised, reproducible interpretation of quantified colours, the reliability of such colour chart can be argued because the interpretation process is subjected to individual perceptions and variations among observers.

More generally, the interpretation of colour based on visual observation is subjective because this method entirely depends on an individual's perception (19, 21). Due to this subjectivity, the ability of individuals to describe colour is inconsistent from time-to-time (23). Hence, it is not surprising that colour descriptions of heat-treated teeth vary from one study to another (1, 10, 20). The lack of objectivity in visual observation method has led to a suggestion to use a quantitative analysis to measure colour changes (21). A study by Rubio *et al.* (24) has evaluated colour of heat-treated teeth using a spectrophotometer. According to their study, the application of spectrophotometry using a Commission Internationale de l'Eclairage lightness (*L\**) and chromaticity (a* and b*) (CIELAB) colour system is reliable as it provides objective colour data with high accuracy. CIELAB was developed by the International Commission on Illumination. Colour values revolve within CIELAB space are lightness (*L\**) and chromaticity (a* and b*). Spectrophotometry also provides tristimulus values (X, Y and Z) that can be used to calculate whiteness and yellowness (luminance) of teeth.

*1.2. Crystallinity analysis*

Crystallinity is an attribute of conformational order within a crystal lattice (25). Teeth exhibit small size of HA crystals around 22 nm with irregularities in the lattice (4). Crystallinity measurement has been used to indicate thermal modification of HA minerals in teeth and bones (25). The growth of HA crystals can be measured using an x-ray diffraction (XRD) analysis, where the output is a diffraction pattern consisting of a set of diffraction peaks. Sharpening of the diffraction peaks in XRD patterns is associated with increased temperature of thermal treatment and increased HA crystallite size (10, 26). Piga *et al.* (4) has indicated the use of XRD to estimate burning temperatures ranging from 200–1000 °C.

*1.3. Aims of the study*

Temperature and duration of heat-exposure are known to influence the degree of structural alterations in teeth including colour, crystallinity and texture (27). This study aimed to quantitatively evaluate effects of temperature and heat-exposure duration on colours and crystallinity of teeth. CIELAB colour space (L*a*b*), luminance (WI and YI), and HA crystallite size were selected as the response variables against a range of temperatures and durations of exposure. We hypothesised that all response variables should change as both temperature and duration change. We also hypothesise that the profile of the observed changes should be the same across all variables. This study also aimed to evaluate the potential of colour and crystallinity of teeth as indicators to estimate temperature of fire.

## 2. Methods

*2.1. Sample preparation*

Mandibular jaws from domestic pigs *(Sus scrofa domesticus)* were obtained from a local abattoir. The pigs were adults of average two years age in which teeth development was completed. Only mandibular bones with fully developed, sound and intact premolars and molars were selected. Mandibles with posterior teeth that had carious lesions were excluded.

Four mandibles were selected. Each mandible was cut into two segments, making a total of eight segments. Each segment had the first premolar as the anterior border and had the second molar as the posterior border. All the attached muscles and fat were completely removed from the bone surfaces using sterile surgical blades. The surgical blades were changed for every mandibular segment. The mandibular segments were cleaned with distilled water. The number of samples are 48 teeth (n = 48). Five teeth from a mandibular segment were removed and kept as control samples at room temperature.

*2.2. Furnace incineration*

The incineration process of the samples was performed in a controlled condition using an electrical furnace (Ward, Serial No: 12098, South Australia) at a laboratory operated by CSIRO Land and Water Division, Urrbrae, South Australia. Mandibular segments were incinerated at each of the following temperature/time combinations: 300°C/30 min, 600°C/30 min, 1000°C/30 min, 300°C/15 min, 600°C/15 min, 800°C/15 min and 1000°C/15 min. Each mandibular segment was placed in a crucible with the buccal surface facing upward. It was then positioned in the centre of the pre-heated furnace for the assigned duration.

*2.3. Visual examination*

Post-incineration, changes in the teeth were examined by visual observation and were recorded as photographic images in the following formatting: Tagged Image File Format (TIFF) and Joint Photographic Experts Group (JPEG). Feature changes including colour, fracture, and any damage seen in teeth were interpreted subjectively.

*2.4. Spectrophotometric analysis*

Colour was measured using a portable handheld spectrophotometer (CM-700d, Konica Minolta Sensing Americas, Inc., U.S.A). An 8-mm target mask with plate was attached at the lens to switch the illumination area. Colour data software (SpectraMagic™ NX CM-S100w) was used to operate the instrument from the computer, record measurements, process the data and for file management.

The spectrophotometer was calibrated twice prior to usage. The first calibration was a zero calibration using a zero calibration box. The second calibration was a white calibration using a white calibration cap with calibration data. The parameters of the spectrophotometer were set to a uniform colour space as recommended by CIE (28), the observer angle was 2° and the illuminant condition was D65. The data collected were the CIELAB L*a*b* values and the luminance values that are X, Y and Z. L* value indicates the lightness of an object's colour on a scale from 0 (black) to 1 (white), and a* and b* are the chromaticity of an object's colour. The value of positive a* is for the redness and negative a* is for greenness. The value of positive b* is for the yellowness and negative b* is for blueness. Measured values of a* and b* near to zero indicates an object of neutral colour, either white or grey.

The whiteness index (WI) and yellowness index (YI) were calculated from the luminance values to analyse the whiteness and yellowness of the teeth. The calculations were performed using the formulae proposed by ASTM Method 313 (29):

$$WI = Y + 800(0.3127 - X) + 1700(0.3290 - Y)$$

$$YI = 100 \frac{[100(1.2985X - 1.1335Z)]}{Y}$$

The buccal surface of the tooth crown was the target area to measure the colour. Five measurements were recorded at the same target area and the average of these measurements were calculated to obtain the final measurement.

*2.5. XRD analysis*

Forty-eight teeth were examined at the X-ray Diffraction Laboratory, CSIRO Land and Water Division, Urrbrae, South Australia, to measure the crystal size of HA. The tooth crowns were removed and manually ground into a powder form in an agate mortar and pestle. Fine powder (~50 mg) were sprinkled onto Si low background holders for XRD analysis. XRD patterns were recorded with a PANalytical X'Pert Pro Multi-purpose Diffractometer using Fe-filtered Co K$\alpha$ radiation, automatic divergence slit, 2° anti-scatter slit and fast X'Celerator Si strip detector. The diffraction patterns were recorded at a scan rate of 2.43°C two theta per minute giving an overall counting time of approximately 30 minutes. Phase identification was performed on the XRD data using in-house software (XPLOT) and HighScore Plus (from PANalytical) search/match software. Calculations of crystallite size were performed using the TOPAS refinement parameter "Cry Size L" that is suggested in the TOPAS-Academic technical reference (30).

*2.6. Statistical analysis*

All statistical analyses were completed using the statistical software R (31). Correlations between response variables were analysed using the built-in 'cor' function. This was necessary to ascertain if multivariate analysis was required. Response variables which are theoretically related and moderately correlated should be fit in a multivariate model. If response variables are weakly correlated than separate univariate analyses can be completed. Alternately, extremely high correlations between response variables suggest redundancy and the number of response variables can be reduced to a representative set.

Linear models were fit to the chosen response variables using the 'lm' function. Possible variable transformations were explored using the boxcox function of the MASS package and all post hoc analyses were completed using a Tukey HSD correction to control the family-wise error rate.

## 3. Results

*3.1. Visual inspection*

Due to the bony protection afforded to the roots, the colour changes primarily focused on the coronal portions of the teeth (Figure 1). For 15-minute heating, teeth colour changed from neutral white-yellowish to opaque white with no signs of crack lines or fracture at 300°C. At 600°C, the tooth crown turned light grey in colour and the crowns begun to noticeably crack.

Visible fracture lines formed through enamel. Premolars became metallic grey and molars turned into black in colour at 800°C, with fragmentation dramatically increased and fracture lines intensified all over the crowns. At 1000°C, the teeth ultimately turned white with complete disintegration of the tooth crowns. There was loss of alveolar bones, making extended fracture lines on the roots visible. For 30-minute heating, the observation contrasts with the 15-minute heating at 600°C where tooth crowns became greyish black and disintegrated whilst the roots were dislodged as alveolar bones cracked. Teeth became chalky white at 1000°C and horizontal fracture lines were noted on the roots.

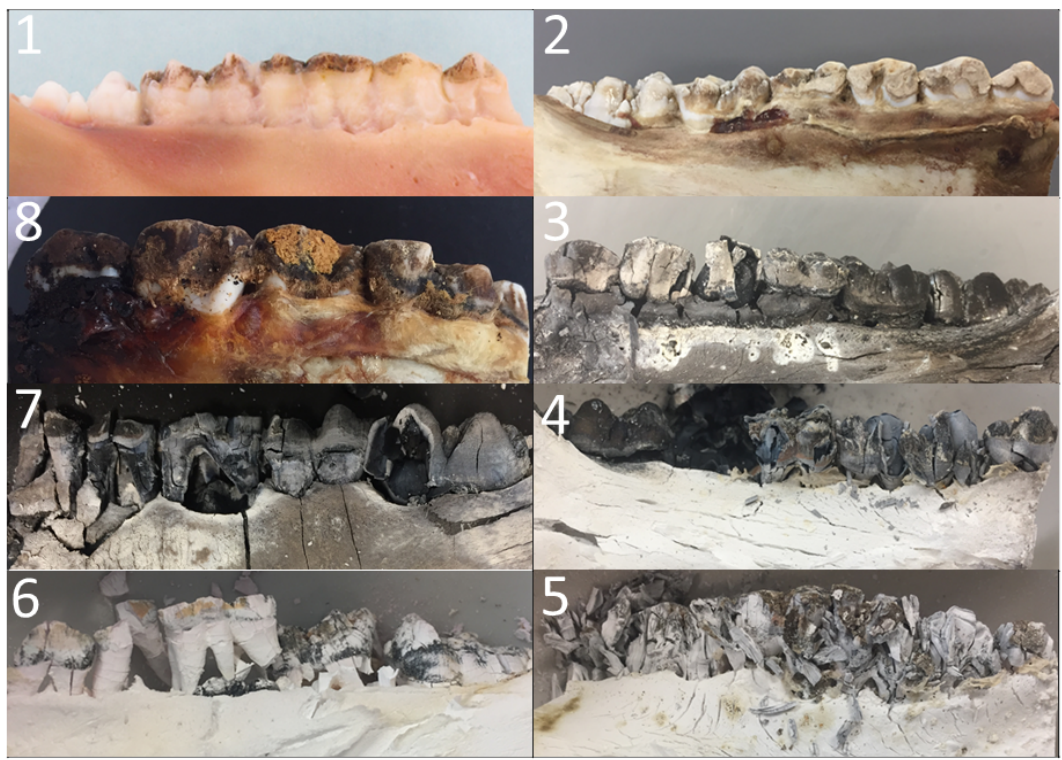

Figure 1—Clockwise from top left: 1) Unheated teeth 2) Teeth heated at 300°C for 15-minute 3) Teeth heated at 600°C for 15-minute 4) Teeth heated at 800°C for 15-minute 5) Teeth heated at 1000°C for 15-minute 6) Teeth heated at 1000°C for 30-minute 7) Teeth heated at 600°C for 30-minute 8) Teeth heated at 300°C for 30-minute.

*3.2. Spectrophotometric data*

The spectrophometric data for the average values of L*, a*, b*, WI and YI of tooth groups at every temperature, separated by duration, are shown in Figure 2. L* values decreased between control group and 800°C. At 1000°C, L* value increased, with the L* value of 15-minute group similar to the control group. L* of the 15-minute group were higher than L* values of 30-minute group at 300° and 1000°C. No apparent changes were observed between chromaticity a* for control teeth with the a* of 15-minute group at 300°C but a twofold-increase of a* for the 30-minute group. At 600°C, a* of the 15- and 30-minute group reduced markedly approaching zero for white colour. The 800°C 15-minute group shows similar values to the 600°C 30-minute group. a* slightly increased from 800°C to 1000°C with a* value of the 30-minute group almost twice the a* value of the 15-minute group. Chromaticity b* increased from the control group to 300°C, with double-hike for 30-minute group and half-increment for 15-minute group, changing for a more saturated yellow colour. A sudden drop of b* value at 600°C for both durations with the yellow of 15-minute group less saturated and the 30-minute group changed towards blue colour. At 800°C, b* was lessened and becoming more neutral yellow. b* was increased by threefold from 800°C to 1000°C, but the yellow saturation for 15- and 30-minute groups were lower than the control group.

WI decreased from the control group to 300°C for both time durations. WI kept on declining till 600°C for the 15-minute group. WI increased for the 600°C 30-minute group and only increased at 800°C for the 15-minute group. However, WI for the 1000 °C 15-minute group rised markedly, in stark contrast to the 30-minute group where the increase was minor. The highest value of YI is manifested in the 300°C 30-minute group before it significantly decreased at 600°C with the lowest value seen in the 30-minute group. A increase of YI is observed from 800° to 1000°C

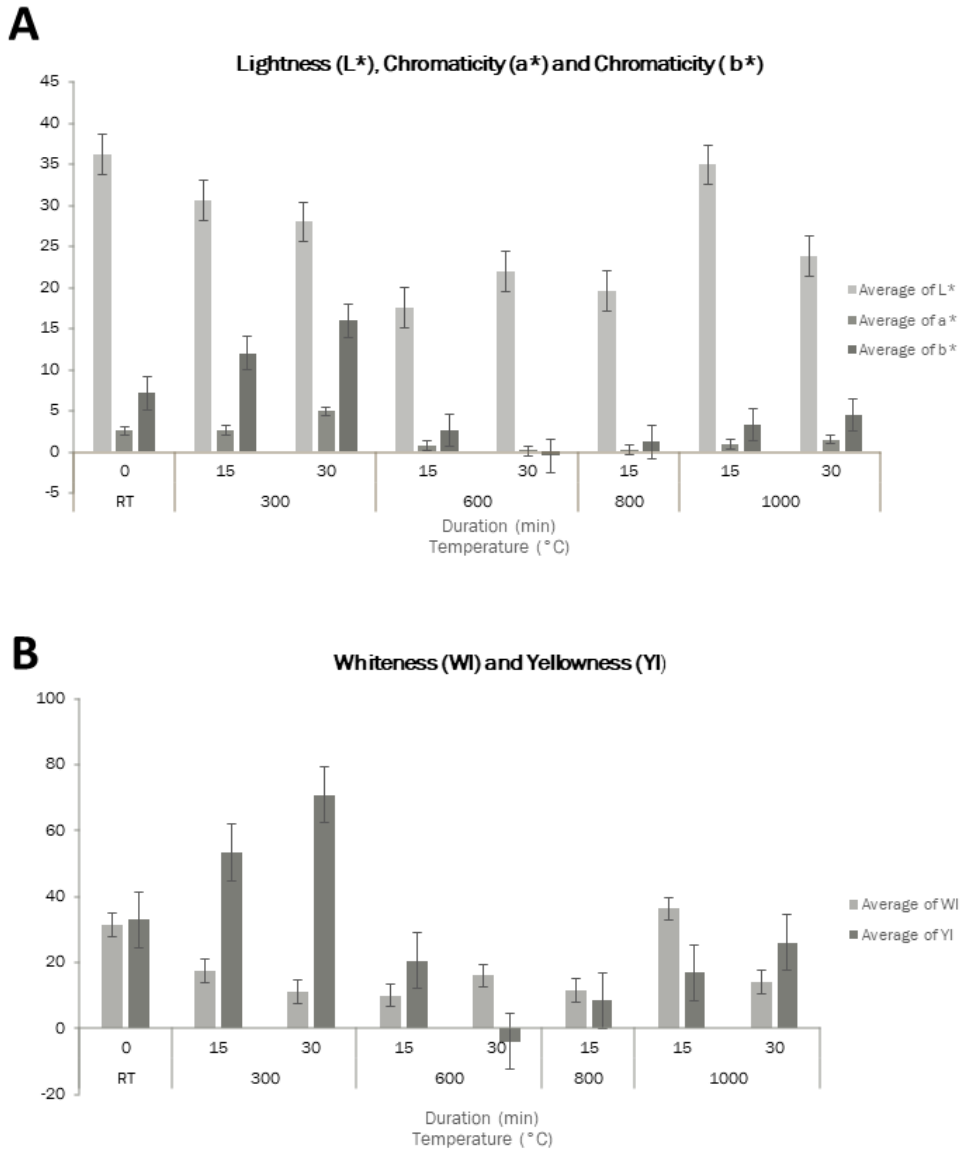

Figure 2—Mean values for (A) lightness (L*), chromaticity (a*) and chromaticity (b*) and (B) whiteness (WI) and yellowness (YI) in tooth groups exposed to room temperature (RT), 300 °, 600 °, 800 ° and 1000 °C for 15- and 30- min.(N = 48).

*3.3. XRD data*

Table 1 lists the average crystallite size of unheated and heated tooth samples. Control samples have an average crystallite size of 63.8nm. Hydroxyapatite crystals gradually shrank when teeth were heated at 300° and 600°C for 15 minutes. However, the 300°C 30-minute group shows dramatic size reduction but a minor decrease in size at 600°C. The crystallite size was then doubled after heating at 800°C for 15 minutes. A dramatic increase, with more than 25-fold augmentation of crystallite size, was observed in teeth heated at 1000ºC for both durations.

Table 1— Average crystallite sizes of hydroxyapatites in teeth left at room temperature (RT) and teeth heated from 300 to 1000°C for 15 and 30 minutes (min) were calculated from (002) peak diffraction pattern. In this study, crystallite size was measured in nanometer (nm). 1 nanometer (nm) = 10 Å (Angstrom)

| Temperature (°C) | Duration (min) | Mean Crystallite Size (nm) |
|---|---|---|
| RT | 0 | 63.8 |
| 300 | 15 | 51.1 |
| 300 | 30 | 39.7 |
| 600 | 15 | 34.4 |
| 600 | 30 | 31.4 |
| 800 | 15 | 73.5 |
| 1000 | 15 | 1940.0 |
| 1000 | 30 | 1999.0 |

Diffraction patterns of control and heated teeth are presented in Figure 3. The diffraction (002) peaks corresponding to hydroxyapatite crystallite and used for crystallite size calculation are marked in Figure 3. In general, the patterns denote a trend concerning the crystallite growth of samples heated for both 15- and 30-minute durations: (i) no notable peak narrowing for control teeth and teeth heated up to 800°C; (ii) Peak narrowed dramatically at 1000°C. Peak widths of teeth heated for 30 minutes appearing similar those heated for 15 minutes. Interestingly, calcite or calcium carbonate ($CaCO_3$) was only detected in samples heated at 300°C for 30 minutes. As expected, the formation of whitlockite was identified in tooth samples heated at 800°C and 1000˚C.

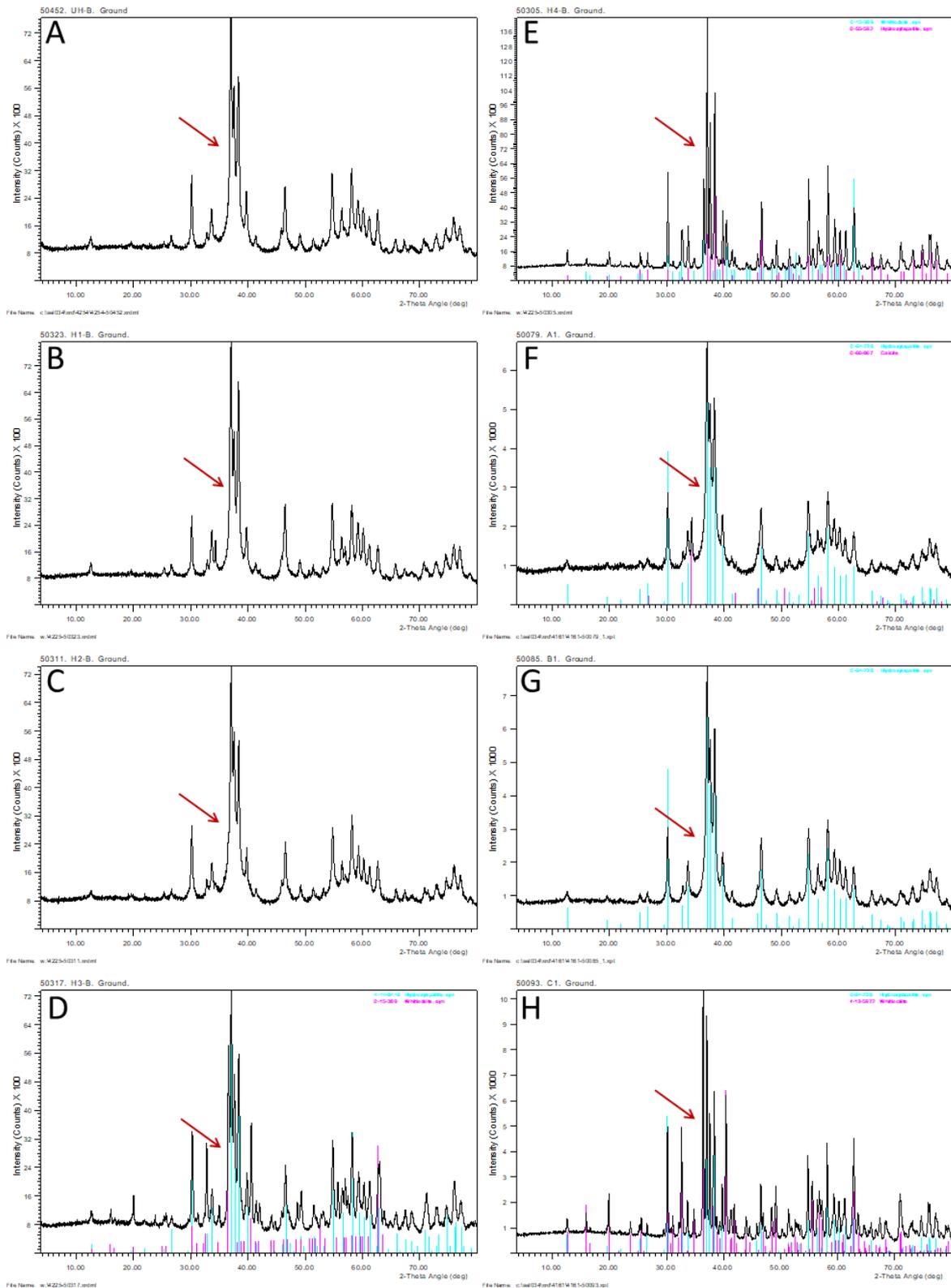

Figure 3—Diffraction patterns of first premolars tooth samples: (A) Unheated (B) Heated at 300°C, 15mins (C) Heated at 600°C, 15mins (D) Heated at 800°C, 15mins (E) Heated at 1000°C, 15mins. (F) Heated at 300°C, 30mins (G) Heated at 600°C, 30mins (H) Heated at 1000°C, 30mins. The diffraction (002) peak of each sample pattern is pointed with an arrow. Sample heated at 300 °C shows the presence of calcite. Samples heated at 800°C and 1000°C for 15- and 30-minutes (A, D, E, and H) reveal the addition of whitlockite.

*3.4. Statistical analysis*

The correlation between response variables was first tested to inform whether multivariate analysis techniques were required. Based on correlation coefficients (for details see Table 2 in supplementary data), strong correlations were found between a* and b* with YI ($p > 0.01$) and between L* and WI ($p > 0.01$), thus a*, b* and L* were omitted from further analysis with YI and WI forming the representative set of response variables. The correlations shown between CS, WI and YI were all weak, justifying the use of separate, univariate analyses rather than a multivariate technique. The predictors for each of these response variables are temperature and duration.

Two-way ANOVAs were then conducted considering both the main effects and interactions between temperature and duration for each of the three response variables, CS, WI and YI. Both CS and WI required log transformations to adhere to the assumptions of homoscedasticity.

*3.5. Crystallite Size (CS)*

Two-way ANOVA revealed a significant effect of temperature on CS ($F(4,31)=199.45$, $p<.001$) and failed to find a significant main effect of duration ($F(1,31)=0.17$, $p=.683$) or interaction between duration and temperature ($F(2,31)=1.25$, $p=.302$). Follow-up post-hoc Tukey HSD pairwise comparisons revealed that the CS of teeth incinerated at 1000°C was between 15.8 and 48.5 higher than the controls, 25.8 and 64.9 higher than 300°C, 35.1 and 88.2 higher than 600°C and 14.1 and 43.2 higher than 800°C (95% confidence intervals), regardless of duration. The only other observable differences were higher CS at 800°C than 600°C (1.3 – 3.9) and higher CS at 600°C than at room temperature (0.3 – 0.9) but these differences are both relatively small compared to those observed at the highest temperature.

*3.6. Whiteness (WI)*

There were no significant main effects or interactions observed for the two-way ANOVA of temperature and duration on WI (Temperature: ($F(4,31)=1.14$, $p=.059$), Duration: ($F(1,31)=0.60$, $p=.597$), Temperature*Duration: ($F(2,31)=2.45$, $p=.103$)). However there was a trend toward significance for the duration variable. Further investigation with a larger data set may be able to explore this difference further.

*3.7. Yellowness (YI)*

The two-way ANOVA of temperature and duration on YI revealed the most complex results in the analysis with a significant interaction between the two predictor variables ($F(2,31)=15.50$, $p<.001$). Pairwise comparisons with a Tukey HSD correction (for details please see Table 3 in supplementary data) found the following:

1. YI peaked at 300 °C. YI was significantly higher at 300 °C than any other temperature, regardless of duration.
2. At 600 °C, YI was between 6.2 and 42.6 higher after 15 minutes, than after 30 minutes. This was the only significant difference based solely on duration.
3. Teeth exposed to 600°C for 30 minutes were between 18.6 and 55.1 higher in YI than the room temperature teeth and between 6.4 and 42.8 higher than teeth exposed to 800°C for 15 minutes. However, teeth incinerated at 600°C for 30 minutes were significantly lower in YI than teeth incinerated at 1000°C for 15 minutes (2.6-39.0) and for 30 minutes (10.6-49.2).

## 4. Discussion

This paper presents findings from spectrophotometric and XRD analyses of heat-treated teeth, and discusses the relationship between colorimetric variables, L*, a*, b*, WI, YI and HA crystal size. Two durations of heat-exposure, 15 and 30 minutes were analysed to investigate the effect of duration on colour changes and crystallite size. Both durations were specifically chosen because previous research has shown colour does not change in teeth after 30 minutes of incineration (20).

### 4.1. *Spectrophotometric analysis*

Despite the visual observation of teeth heated for 15-minutes and 30-minutes being different, our spectrophotometric results were generally unable to find a significant effect of duration on colour variables or crystal size ($p < 0.05$). This finding is in stark contrast with a previous study in which duration of heat-exposure was shown to affect tooth colour changes (21). Our results also showed that the lightness, L* value, is not affected by temperature. This finding contradicts the findings in a study by Rubio *et al.* (24) that L* is a valuable determinant to estimate the incineration temperature. Our findings are interesting considering that dehydration was shown to have significant effect on the lightness of teeth (32). However, the dehydration process is only accountable for the heating up to 500°C (8, 10). In our study, the measure of lightness decreased from the room temperature to 600°C, a range where teeth were losing the water component due to dehydration. Enamel was mainly affected as it is entirely made up of inorganic components and therefore, it became progressively porous. This finding is supported by studies in which the lightness of tooth was found to be proportionate porosity of enamel and decreased when they are dehydrated (33, 34). In addition, a major advantage of using the CIELAB colour system is that the difference between a standard and the measured colour can be quantified (35).

4.2. *XRD analysis*

In general, the most significant transformation of HA crystallite phase occurs after 800°C. Reduction in crystallite size in 300°C to 600°C sample groups could be due to the dehydration process that altered the chemical structure of HA. By 800°C, HA crystallites progressively expanded to fill up the interspaces resulted from the combustion of organic matters. The sudden and striking growth of the crystallite size noted in teeth heated at 1000°C is consistent to a previous study by Piga *et. al.* (4).

The observed narrowing of the diffraction peaks is related to changes in the crystallite size that increases with heating temperature. Unheated teeth show relatively broader peaks than heated which reflect a poor crystalline apatite. Diffraction peaks became narrower and sharper, corresponding to an increase in the crystal size of hydroxyapatite and towards a more orderly crystalline arrangement in lattice plane, are observed as the temperature increased up to 1000°C. These results are consistent with the most significant structural changes of the bone mineral occurring between 600°C and 800°C, as previously reported (10, 25, 36).

Interestingly, calcite only appeared in samples heated to 300°C for 30-minute. No calcite was detected in specimens of the 300°C 15-minute group. It was also absent in specimens of the 600°C group. During heat treatment, calcium carbonate is formed when hydroxyl ions ($OH^-$) of hydroxyapatite are replaced by carbonate ions $CO_3$ (37). This might explain why there is a marked decrease in size for the 300°C 30-minute group. Calcite is the most stable form of calcium carbonate ($CaCO_3$) (38). It was reported that in bone, calcium carbonate formed after a complete combustion of organic matter around 600°C and it disappeared after heating at 900°C (39). Contrasting to that, other studies have claimed that complete loss organic matters was at 350°C, with carbonate ions decreasing around 400-700°C (40, 41). By far, findings on the temperature at which calcite is formed and loss are greatly varied and mostly inconsistent. Being the third most abundant component of skeletal minerals (42), accidental carbonate contamination by adsorption on the surface of apatite crystals cannot be ruled out. The mechanism of calcite formation exclusively in the 300°C 30-minute samples is still obscure. Thus, further studies on the occurrence of calcite in burned dental remains are needed.

We found that at temperatures over 800°C whitlockite formed. Whitlockite is a product of thermally-heated hydroxyapatite where over 800°C, HA crystals coalesce to each other (43, 44). This finding corresponds to the whitlockite seen on the XRD pattern in human teeth heated after 750˚C (45, 46). The claim that no definite temperature at which transformation of bioapatite to whitlockite occur is therefore arguable (45). The evidence of hydroxyapatite becoming whitlockite can aid in the estimation of temperature. The results presented here suggests that identification of whitlockite in burned teeth means the teeth have been heated at no less than 750˚C. This is an important indicator since the temperature of house fires, motor-vehicle accidents and fire disasters are 750°C and above (7, 47). Also, whitlockite was consistently identified in all teeth heated to more than 800°C, a finding that supports a study by Piga *et al.* of which whitlockite formed systematically in teeth heated at 750°C (45).

There was, however, a dramatic increase in crystallite size at 1000˚C suggesting that the HA crystallites may have recrystallised and expanded drastically to fill the voids created by dehydration and organic matter destruction. Dehydration or loss of water in the enamel and dentin occur in between 100 – 400˚C where hydroxyl bonds break during the process and water molecules evaporated, followed by the loss of organic matter after 400°C (9). Sudden increase of HA crystal size is noticeably consistent for all teeth heated at 1000˚C. This finding could be a useful feature to indicate if a dental remain has been subjected to temperatures higher than 800˚C. Our results correspond to a previous study where a marked increase in crystallinity with more ordered crystal lattice has been found in teeth heated above 700°C (4).

4.3. *Recommendations*

We recommend the integration of spectrophotometric and XRD analyses on burned teeth to estimate the maximum heating temperature with greater accuracy. The credibility of spectrophotometric and XRD analyses when used individually to recognise burned skeletal remains and to estimate the maximum heating temperature have been proven in some studies (10, 48, 49). Yet, the conclusions from the individual use of spectrophotometric or XRD analysis alone provide weak evidence of incineration for three reasons. Firstly, any change of colour seen in tooth does not necessarily means that it has been burned. The colour change can be influenced by multiple factors such diet, nicotine staining or fluorosis (19). Secondly, the growth of hydroxyapatite increased in a diagenetically altered teeth as seen in bone (50). Diagenesis is a process that starts after the vertebrates die in which taphonomic factors such as elements of burial soil and the presence microorganisms influenced the alteration of the chemical structure and induced recrystallization in teeth and bones (51, 52). The increased of crystallinity was found to be inversely correlated with the collagen contents (53). Thirdly, as an increase of crystal size is notable only from teeth heated more than 700 °C, the sole use of XRD is not very useful for estimating low temperatures.

Since the application of spectrophotometric and XRD analyses on real archaeological or forensic scenarios are still uncommon, we also recommend both analyses to be incorporated in the future research investigating heat-treated teeth. Combining analytical techniques to investigate burned skeletal remains has been suggested in a previous archaeology research to improve the robustness and accuracy of the interpretation (54). Measuring the colour and crystallinity of heat-treated teeth using both analyses have enabled us to obtain quantitative data with precision and accuracy. Both analyses are also cost-effective, require small amount of sample (0.5 mg) and the procedure is relatively simple. Indeed, instrument analyses offers objective, quantifiable, repeatable and quick measurement readings (55). In this study, the spectrophotometer was calibrated to a white reference prior to measuring the teeth. The white calibration step allows the normalisation of the spectrophotometer to correctly obtain correct transmittance or reflectance factors throughout spectrum (56). The application of spectrophotometer also can minimise interpretation errors (57). XRD on the other hand has been claimed to be the ideal mean to express the crystallinity of a bioinorganic phase (45).

## 5. Conclusion

The longevity of burned dental remains has greatly benefited studies of archaeological and forensic sciences. This study recommends the integration of spectrophotometry and XRD analyses to be a routine practice in casework upon the discovery of burned dental remains. Simultaneous used of both analyses has proved to be able to accurately estimate maximum heating temperature of teeth. Also, the inclusion of both analyses in the investigation of burned dental remains could potentially amplify the strength of evidence in archaeological and forensic casework.

**Acknowledgments**

This work has been possible thanks to Professor Robert Fitzpatrick (Acid Sulfate Soils Centre (ASSC) | Centre for Australian Forensic Soil Science (CAFSS), The University of Adelaide) for providing expertise, facility and equipment at Water and Land division, CSIRO, Waite,

South Australia. The authors acknowledge the financial support received for this research from the School of Biological Sciences, the University of Adelaide.**Conflict of interest**

The authors declare no conflict of interest with this research

**Ethics**

The handling of the animal remains in this study was done in line with the University of Adelaide Animal Ethics.

**References**

1. Karkhanis S, Ball J, Franklin D. Macroscopic and microscopic changes in incinerated deciduous teeth. The Journal of forensic odonto-stomatology. 2009;27(2):9-19.
2. Muller M, Berytrand MF, Quatrehomme G, Bolla M, Rocca JP. Macroscopic and microscopic aspects of incinerated teeth. J Forensic Odonto-Stomatol. 1998;16(1):1-7.
3. Sui T, Sandholzer MA, Le Bourhis E, Baimpas N, Landini G, Korsunsky AM. Structure-mechanical function relations at nano-scale in heat-affected human dental tissue. Journal of the Mechanical Behavior of Biomedical Materials. 2014;32:113-24.
4. Piga G, Thompson TJ, Malgosa A, Enzo S. The potential of X-ray diffraction in the analysis of burned remains from forensic contexts. J Forensic Sci. 2009;54(3):534-9.
5. Higgins D, Austin JJ. Teeth as a source of DNA for forensic identification of human remains: A Review. Sci Justice. 2013;53(4):433-41.
6. Robinson F, Rueggeberg F, Lockwood P. Thermal Stability of Direct Dental Esthetic Restorative Materials at Elevated Temperatures. J FORENSIC SCI. 1998;43(6):1163-7.
7. Fairgrieve SI. Forensic Cremation: Recovery and Analysis. Forensic Examiner. 2009;18(1):74-5.
8. Schmidt CW. Chapter 3 - Burned Human teeth. In: Schmidt CW, Symes SA, editors. The Analysis of Burned Human Remains (Second Edition). San Diego: Academic Press; 2015. p. 61-81.
9. Driessens FCM, Verbeeck RMH. Biominerals Ferdinand C. M Driessens, Verbeeck RMH, editors. Boca Raton CRC Press; 1990. 428 p.
10. Shipman P, Foster G, Schoeninger M. Burnt bones and teeth: an experimental study of color, morphology, crystal structure and shrinkage. J Archaeol Sci. 1984;11(4):307-25.
11. Kubisz L, Mielcarek S. Differential scanning calorimetry and temperature dependence of electric conductivity in studies on denaturation process of bone collagen. Journal of Non-Crystalline Solids. 2005;351(33):2935-9.
12. Ramachandran GN. Stereochemistry of collagen*. International Journal of Peptide and Protein Research. 1988;31(1):1-16.
13. Thompson TJU. Heat-induced dimensional changes in bone and their consequences for forensic anthropology. Journal of Forensic Sciences. 2005;50(5):1008-15.
14. Sakae T, Mishima H, Kozawa Y, Legeros RZ. Thermal stability of mineralized and demineralized dentin: A differential scanning calorimetric study. Connective Tissue Research. 1995;33(1-3):193-6.
15. Vargas-Becerril N, Reyes-Gasga J, García-García R. Evaluation of crystalline indexes obtained through infrared spectroscopy and x-ray diffraction in thermally treated human tooth samples. Materials Science and Engineering: C. 2019;97:644-9.
16. Jang HL, Jin K, Lee J, Kim Y, Nahm SH, Hong KS, et al. Revisiting whitlockite, the second most abundant biomineral in bone: Nanocrystal synthesis in physiologically relevant conditions and biocompatibility evaluation. ACS Nano. 2014;8(1):634-41.18